# Social Science Theories in Software Engineering Research

Tobias Lorey
Department of Computer Science,
University of Innsbruck
Innsbruck, Austria
tobias.lorey@student.uibk.ac.at

Paul Ralph
Faculty of Computer Science,
Dalhousie University
Halifax, Canada
paulralph@dal.ca

Michael Felderer
Department of Computer Science,
University of Innsbruck
Innsbruck, Austria
michael.felderer@uibk.ac.at

## ABSTRACT

As software engineering research becomes more concerned with the psychological, sociological and managerial aspects of software development, relevant theories from reference disciplines are increasingly important for understanding the field's core phenomena of interest. However, the degree to which software engineering research draws on relevant social sciences remains unclear. This study therefore investigates the use of social science theories in five influential software engineering journals over 13 years. It analyzes not only the extent of theory use but also what, how and where these theories are used. While 87 different theories are used, less than two percent of papers use a social science theory, most theories are used in only one paper, most social sciences are ignored, and the theories are rarely tested for applicability to software engineering contexts. Ignoring relevant social science theories may (1) undermine the community's ability to generate, elaborate and maintain a cumulative body of knowledge; and (2) lead to oversimplified models of software engineering phenomena. More attention to theory is needed for software engineering to mature as a scientific discipline.

## CCS CONCEPTS

• **General and reference** → **Surveys and overviews**; • **Software and its engineering**;

## KEYWORDS

software engineering, theory, social science

## 1 INTRODUCTION

The software engineering (SE) research community increasingly accepts that software projects have important psychological, sociological and managerial properties [59]. SE research has therefore increasingly considered software development's human and organizational aspects alongside its formal and technical aspects [31]. Embracing the social aspects of software development involves incorporating theories and research methods from social science—including psychology, sociology and management (especially information systems)—which study individuals, groups and organizations.

On one hand, *methods* from social sciences (e.g. case studies, experiments, questionnaires), which investigate individuals, groups or organizations, are now widely used in software engineering [82]. On the other hand, *theories* from social sciences—systems of ideas for explaining, describing, predicting or analyzing human phenomena [26]—do not seem widely applied in SE research despite being critical to advance SE as a scientific discipline. One reason could be SE's formal and technical roots [72], while social phenomena were historically marginalized [30].

However, researchers are increasingly recognizing that stronger theoretical foundations are needed for SE to mature as an applied science [70]. Several initiatives for increasing attention to theory in SE have been proposed. Sjoberg et al. [68] devised a framework for describing theories and guidelines for proposing, employing, testing and modifying them. Hannay et al. [27] conducted a systematic literature review of theories used in software engineering experiments. SE-specific guidelines have been created for theory development [58, 73]. General theories of software engineering have been proposed [31, 56]. Stol & Fitzgerald [70] argued for training PhD students in theorizing. Finally, a series of international workshops explored possibilities for general and core theories in SE [21, 33, 60–62].

This paper extends these initiatives with a focus on theories from social sciences. It aims to increase attention to social science theories and help SE researchers use social science theories more effectively. Increasing attention to social science theories will help researchers to:

(1) create, accumulate and preserve knowledge [62, 68, 70];
(2) cope with the multi-disciplinarity of SE research [71];
(3) improve the quality of their research [70]; and
(4) reduce the need for trial and error [32, 62].

More generally, "theory provides explanations and understanding in terms of basic concepts and underlying mechanisms, which constitute an important counterpart to knowledge of passing trends" [27]. Social science theories help us understand core SE phenomena (e.g. programming) and their impacts on individuals, teams, projects and organizations [68].

This paper therefore analyzes which, how, where, and to what extent social science theories are used in SE research. It provides a complete list of the theories encountered, elaborating on the most common. It explains how inattention to social science leads SE researchers to oversimplify and over-rationalize core phenomena. Finally, it highlights areas in which social science theories may be especially relevant, but are presently neglected.

This paper is structured as follows: Section 2 presents relevant background and related work on theories in social sciences and SE, including an overview of the classification criteria employed by this paper. Sections 3 and 4 describe the research method and present the results. Section 5 elaborates on the theories used most often in SE research. Section 6 discusses the study's implications, limitations and avenues for future research. Finally, Section 7 concludes the paper with a summary of its contributions.

## 2 BACKGROUND

This section explains our operationalization of social science and the classification schemes we later employ.

We consider a theory a social science theory if it originates in one of the disciplines listed by *The Social Science Encyclopedia* [36]: anthropology, cognitive science, criminology and law, cultural studies, demography, economics, education, evolution, gender, geography, health and medicine, history, industrial relations and management, language, linguistics and semiotics, mental health, methods of social research, philosophy, political theory, politics and government, psychology, social problems and social welfare, and sociology. Some notes on this list are warranted:

- Some of these fields (e.g. geography) mix social, natural or applied science.
- Some fields intersect several broader social sciences (e.g. communication studies intersects cultural studies, linguistics, management, and sociology).
- Each field has many subfields (e.g. information systems, which shares many similarities with SE, is a subfield of management).
- We are not sure that philosophy is a social science, per se, but tentatively include it anyway.

Theories can be classified on numerous dimensions including what they are for, their scope and their structure, as well as the way they are used and where in an article they are used.

Gregor [26] proposed a taxonomy for classifying information systems theories according to their purpose. She argued that theories can serve four purposes (analysis and description, explanation, prediction, and prescription), leading to five types:

1. *Theories for analyzing*, also called frameworks or taxonomies, are relatively simple, descriptive theories. A good example is Iivari et al.'s dynamic four-tiered framework for classify information systems development methodologies and approaches [28].
2. *Theories for explaining*, also called *theories for understanding*, attempt to explain what, how, when, why or where phenomena occur. They do not include causal hypotheses or make numerical predictions. Darwinian evolution is a theory for explaining, as is dual coding theory [49].
3. *Theories for predicting* predict future events without specifying causal relationships (e.g. weather forecasting models). These theories have testable propositions but no prescriptions.
4. *Theories for explaining and predicting* posit causal explanations and testable propositions (e.g. quantitative hypotheses). Prescriptions may be included but are not the focus. This type is common in information systems, where examples include the technology acceptance model [17], cognitive fit theory [81], and the theory of task-technology fit [25].
5. *Theories for design and action* provide detailed instructions on how to do something, like recipes. These are less common in the social sciences than in software engineering. Multiview, the methodology and contingency framework, is a good example [3].

Theories also exist on a kind of generalizeability spectrum from observations and empirical generalizations to middle-range theories, general theories, grand theories, and theories of everything [61, 72]. SE researchers are mainly concerned with empirical generalizations, middle-range theories and general theories [72]. An *empirical generalization* is "an isolated proposition summarizing observed uniformities of relationships between two or more variables" [44]; for instance, Moore's law. *Middle-range theories* (e.g. COCOMO [8]; the software engineering principles of Davis [16]) "lie between the minor but necessary working hypothesis that evolve in abundance during day-to-day research and the all-inclusive systematic efforts to develop a unified theory that will explain all the observed uniformities" [44]. A *general theory*, meanwhile, "applies across [a] field and unifies existing empirical and theoretical work" [33]. No general theory of software engineering is widely accepted, but a few have been proposed (e.g. the tarpit [31]; sensemaking-coevolution-implementation theory [55]).

Researchers also distinguish between variance theories, process theories, and taxonomies. Variance theories explain and predict the *variance* in one or more dependent variables using independent, moderating and mediating variables [14]. Process theories explain how an entity changes and develops [78]. Taxonomies (also called theories for understanding, descriptive theories, frameworks and typologies) organize instances (individual things) into classes (abstract descriptions) [58].

Variance theories typically posit causal relationships among constructs [63]; process theories and taxonomies typically do not [58]. Process theories usually focus on non-causal relationships between entities (e.g. actors, objects) while taxonomies focus on similarities between entities. Variance theories are often associated with positivist, quantitative, statistical research while process theories and taxonomies are often associated with interpretivist, qualitative, non-statistical research [45]. In many scientific disciplines, including information systems, variance theories are more common than process theories. However, process theories better explain the relationships between inputs and outcomes [14]. Variance theories and process theories can sometimes be combined [48].

Meanhile, theories can serve different purposes within a research article [27]:

- *Design:* The theory influences a study's research method including research questions, hypotheses, tasks, materials, simulation parameters, or research models and frameworks.
- *Explanation:* The theory explains a study's results. There may be a fine line between a theory that explains results and a theory that analyzes results.
- *Applied:* The theory is used for its intended purpose (e.g. predicting technology adoption) within a study, or to analyze study results.
- *Motivation:* The theory inspires the article's topic (but its concepts are not used explicitly to formulate the study design; that would be *Design*).

- *Tested:* The article evaluates the theory using primary data (e.g. an experiment) or secondary data (e.g. a systematic literature review).
- *Modified:* The article incrementally modifies, adapts or extends the theory.
- *Basis:* The article proposes a new theory, with its own character and concepts, partly based on an existing theory.

Hannay et al. [27] include an eighth role—*proposed*—however, theories proposed by SE articles are beyond the scope of this review because they do not originate in a social science. We added the role *applied* to accommodate theories that are used for their intended purpose. Furthermore, we follow Hannay et al. [27] in excluding tangential references to theories and descriptions of theories, as in a related work section, and focus on *substantial* uses of theories.

Finally, we can classify theories according to where in an article they are used. The IMRaD pattern—"Introduction, Method, Result, and Discussion"—is the de-facto standard for scientific writing in many fields [6]. (In IMRaD, *introduction* includes related work while *discussion* includes implications, limitations, future work and conclusions.) While each article is unique, most can be mapped to the IMRaD structure. We can ignore front and back matter such as the title, abstract, keywords, references, acknowledgements, and appendices as these sections rarely contain substantive uses of theories.

## 3 METHODOLOGY

We conducted a critical review. Critical reviews [e.g. 4, 73, 83] are similar in execution to systematic reviews [cf. 34], except that critical reviews analyze a sample of papers to make a point (often about theory or methods), whereas systematic reviews aim to synthesize all relevant evidence.

Briefly, we identified clear and upfront analysis questions, selected a search strategy including explicit inclusion and exclusion criteria, retrieved a sample of articles from prominent SE journals, extracting instances of theory use from the articles, and analyzing the theories and their uses by applying a prior coding scheme.

### 3.1 Objective and Questions

This study aims to provide a comprehensive summary of the state of social science theory use in software engineering research. We refine this goal into five specific analysis questions (AQs) all of which end with an implied 'in SE research':

(AQ1) To what extent are social science theories used?
(AQ2) What social science theories are used?
(AQ3) Which social sciences provide the theories?
(AQ4) Which types of theories are used?
(AQ5) For which purposes are theories used?
(AQ6) Where in the papers are theories used?
(AQ7) In what knowledge areas are theories used?

### 3.2 Journal, Article and Theory Selection

We selected five SE journals to study: *IEEE Transactions on Software Engineering (TSE), ACM Transactions on Software Engineering and Methodology (TOSEM), Empirical Software Engineering (EMSE), the Journal of Systems and Software (JSS),* and *Information and Software Technology (IST).* We chose these journals because they are widely considered to be among the most prestigious, influential and academically rigorous journals that sit squarely in SE.

To keep the review recent, tractable and reproducible, we limit our search to articles published between 1 January 2007 and 31 December 2019. Within this thirteen-year period, the five selected journals published more than 5,900 articles. Of these, we manually excluded editorials, introductions, book reviews, and other non-research articles.

We scanned the remaining research articles for theory use according to the following process:

(1) We performed an electronic search for the term "theory" within each entire article, which can serve as an initial indicator for theory use.
(2) An electronic search for the names of the most commonly used theories in information systems according to the Theories Used in IS Research Wiki[1] or Dwivedi et al. [19]. This step was included to avoid missing theories that do not have the word "theory" in the title (e.g. *Task Technology Fit*) and potentially explore similarities between SE and IS.

Each paper returned by these searches was then manually checked against the following inclusion criteria:

(1) the "theory" meets our definition of a collection of ideas for understanding, analyzing, explaining or predicting;
(2) the theory takes on one of Hannay et al.'s roles as adapted for our study (Section 2);
(3) the theory originates in one of the social science subfields listed in the Social Science Encyclopedia [36]. (If not obvious, we examined the reference list or searched for the theory in relevant scientific databases).

We exclude theories that originated in software engineering, mere observations, grand theories, theories of everything, and theories that were mentioned but not substantively used (Fig. 1).

### 3.3 Data Extraction and Analysis

Next, we extracted relevant data from the articles. We extracted metadata including title, authors, journal, volume, issue (if applicable) and publication. In addition to metadata, we extracted data on the theories and articles themselves. For theories, we extracted the originating discipline, theory role, theory usage, theory nature, and Gregor's theory type. For articles, we extracted the corresponding SWEBOK classification.

To gather this data, we analyzed each *theory* to determine its originating discipline, type (according to Gregor [26]), scope (middle-range, etc.) and nature (process, variance or taxonomy). Then we analyzed each *theory usage* to determine the theory's role and where it was used (introduction, methods, results or discussion).

Finally, we classified each article according to the fourteen knowledge areas defined by the software engineering body of knowledge (SWEBOK) [10]—design, testing, etc.

*3.3.1 Edge Cases.* This paper focuses on theories imported from social sciences. We therefore exclude theories from mathematics (e.g. chaos theory), natural sciences, (e.g. the theory of evolution) and theories homegrown in SE (e.g. theory-w [9], sensemaking-coevolution-implementation theory [55], the theory of distances in SE [7]). Of course these other theories are also important—they are just not the focus of this study.

---
[1]https://is.theorizeit.org/

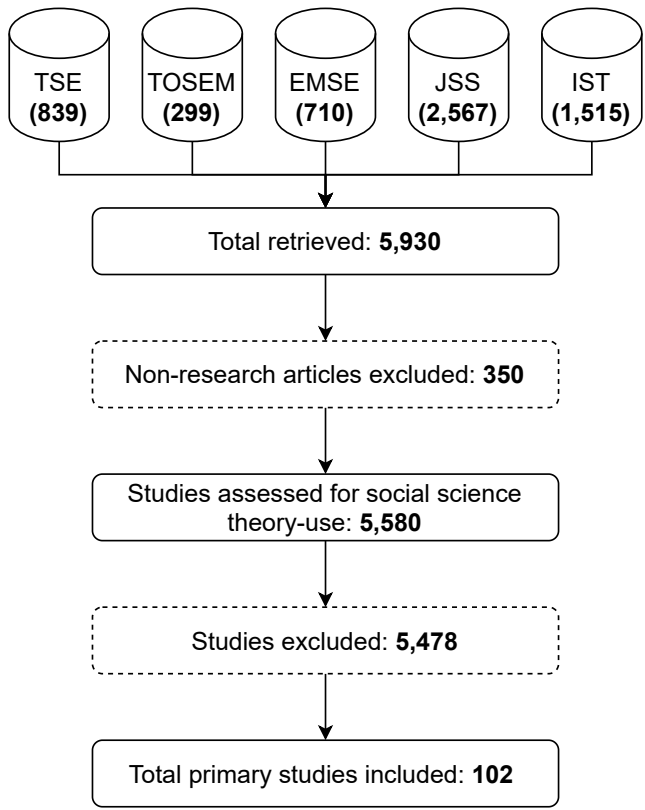

Figure 1: PRISMA Diagram

We also excluded SERVQUAL [50] because although it is often presented as a theory, it is clearly an *instrument* [75]. We do, however, include expectation disconfirmation theory, from which SERVQUAL is derived.[2]

Three edge cases we did include are coordination theory, real options theory and argumentation theory. Coordination theory and real options theory could be seen as a bodies of research (like game theory). However, we included coordination theory because, in some usages, it clearly adapts sociological concepts to software engineering. Meanwhile, we conceptualize real options theory as a theory for analyzing (Gregor's Type I) because it is used to analyze investment decisions. Toulmin's [77] argumentation theory, in contrast, is sometimes seen as a method; however, we again consider it a Type I theory because it facilitates analyzing the structure of an argument.

### 3.4 Validation

We validated our analysis in two ways. First, the second and third authors performed a detailed audit of the first author's classifications. This review produced the edge case decisions described in

---
[2]SERVQUAL is typically capitalized despite being a portmanteau of 'service' and 'quality', not an acronym

Section 3.3.1 and many improvements to the classification of theories. Next, we extracted the authors' contact information from each article that applied one or more theories and sent the authors a questionnaire regarding their article, the theory (or theories) they used and their experience with social science theories. Questionnaire results are reported in Section 4.8

## 4 RESULTS

This section presents our results, organized by analysis question. Below, we reference primary studies as *[P...]*. A comprehensive replication package, including the complete list of primary studies, is available (see Data Availability).

### 4.1 To what extent are social science theories used in SE research?

Only 102 (1.8%) of the 5580 included articles used a social science theory in some substantial way. Only 24 used more than one theory. For example, Moody [P1] justifies scientific foundations for creating visual notations using six theories: the Shannon-Weaver model, cognitive fit, cognitive load, dual coding, feature integration theory, and the theory of symbols. Meanwhile, Prechelt and Oezbek [P2] used the garbage can model, actor-network theory, structuration theory, and path dependence to study open source software process innovation. More recently, Menolli et al. [P3] employed concepts from organizational learning, the theory of communities of practice, and Nonaka's dynamic theory of organizational knowledge creation to knowledge sharing and learning in software development companies.

### 4.2 What social science theories are used in SE research?

The primary studies substantially used 87 unique theories. Of these, 27 (31%) appear in two or more articles (Table 1) while 60 are used in a single article (Table 2). The most commonly used theory is the technology acceptance model (TAM) [17]. Tied for second place, with five uses each, are dual coding theory and diffusion of innovations.

### 4.3 Which social sciences provide the theories used in SE research?

The 87 theories used originate from eight social sciences: psychology (34), industrial relations / management including information systems and organizational studies (23), economics (10), communication studies (8), sociology (8), education (2), linguistics / semiotics (1), and philosophy (1). (Communication studies was not included in the Social Science Encyclopedia because it is an "intersecting discipline"—see Section 2— but the eight theories in question are best understood as originating in the interdisciplinary communication studies community.)

Table 1: Overview of theories used in more than one article

| Theory (T.) | Ref | Description | From[1] | n | Articles |
|---|---|---|---|---|---|
| Technology Acceptance Model | [17] | posits that usefulness and ease-of-use cause intention to adopt | MIS | 13 | [P4-P15, P20] |
| Diffusion of Innovations T. | [65] | explains how ideas and products (innovations) spread over time | Comm. | 5 | [P16-P20] |
| Dual Coding T. | [49] | posits that verbal and nonverbal stimuli are processed by different cognitive subsystems | Psyc. | 5 | [P1, P21-P24] |
| Coordination T. | [41] | a body of principles about how the activities of separate actors can be coordinated | Org. | 4 | [P25-28] |
| T. of Reasoned Action | [23] | posits that attitude and subjective norm lead to behavioral intention and actual behavior | Psyc. | 4 | [P20, P39-P41] |
| Cognitive Fit T. | [81] | posits that problem solving performance depends on the alignment between problem representation and tasks | MIS | 3 | [P1, P23, P29] |
| Cognitive T. of Multimedia Learning | [43] | explains how multimedia representations are processed, remembered and recalled | Psyc. | 3 | [P22, P30, P31] |
| Organizational Learning T. | [2] | explains the processes by which organizations create knowledge, including single- and double-loop learning | Mgmt | 3 | [P3, P27, P32] |
| Real Options T. | [46] | analyzes the value of potential investments by modeling them as options (like stock options) | Econ. | 3 | [P33-35] |
| Media-Richness T. | [15] | posits that richer communication mediums improve communication effectiveness | Comm. | 3 | [P36-P38] |
| Cognitive Load T. | [74] | explains how cognitive load affects learning | Psyc. | 3 | [P1, P42, P43] |
| Task-Technology Fit | [25] | posits that the alignment between technologies and tasks improves performance and utilization | MIS | 3 | [P7, P37, P49] |
| Organizational Knowledge Creation | [47] | explains how an organization's knowledge is created by a dialogue between tacit and explicit knowledge | Org. | 2 | [P3, P44] |
| Structuration T. | [24] | understands human behavior in terms of interactions between structure and agents | Soc. | 2 | [P2, P45] |
| Control T. | [35] | explains how control is carried out in organizations to make employees adhere to certain standards | Org. | 2 | [P46, P47] |
| Agency T. | [29] | explains the relationship between agents and principals where agents make decisions on behalf on principals | Org. | 2 | [P46, P48] |
| Information Foraging T. | [52] | explains human information search mechanisms and posits that people have built-in search mechanisms | Psyc. | 2 | [P50, P51] |
| Goal-Setting T. | [39] | explains how goals affect task performance; posits that motivating goals improve performance | Psyc. | 2 | [P52, P53] |
| Contingency T. | [22] | posits that organizational outcomes depend on the alignment between the leadership style and the situation. | Org. | 2 | [P54, P55] |
| T. of Argumentation | [77] | describes how humans argue and how arguments are structured | Comm. | 2 | [P56, P57] |
| Activity T. | [20] | describes activities of subjects in relation to the community and other constructs | Psyc. | 2 | [P58, P59] |
| Transaction Cost Econ. | [13] | explains how costs occur when transactions are performed within a market | Econ. | 2 | [P60, P61] |
| UTAUT | [79] | explains why organizations adopt technologies | MIS | 2 | [P62, P63] |
| Human Error T. | [64] | describes a four stage classification of how humans produce errors | Psyc. | 2 | [P64, P65] |
| Contact Hypothesis | [1] | explains mechanisms of group contact that reduce prejudice | Psyc. | 2 | [P66, P67] |
| Resource Dependency T. | [51] | studies the relationship between an organization and its resources | Soc. | 2 | [P61, P53] |
| Media Synchronicity T. | [18] | examines media in the context of synchronicity | MIS | 2 | [P37, P72] |

[1]Comm.=Communication Studies; Econ.=Economics; MIS=Management Information Systems; Mgmt=Management; Org.=Organizational Studies; Psyc.=Psychology; Soc.=Sociology

Table 2: List of theories used in one article

| | | | |
|---|---|---|---|
| Activity t. | Cultural lag t. | Personal construct t. | Subjective t. of value |
| Actor-network t. | Cumulative prospect t. | Power dependency t. | Task-media fit |
| Adaptive structuration t. | Dickinson and Mcintyre's teamwork model | Prejudice t. | The integrated model of group development |
| Bases of power | | Rhetorical t. of diffusion | T. of affordances |
| Bounded rationality | Dual-process t. of cognition | Shannon-Weaver model | T. of argumentation |
| Cognitive apprenticeship t. | Dynamic capability t. | Shared mental models t. | T. of group productivity |
| Cognitive-based view | Expectancy t. | Situated cognition t. | T. of intellectual capital |
| Cognitive model of media choice | Expectation disconfirmation t. | Social capital t. | T. of planned behavior |
| | Experiential learning t. | Social cognitive t. | T. of power and conflict |
| Common ground t. | Feature integration t. | Social education t. | T. of symbols |
| Common teamwork t. | Garbage can model | Social identity t. | T. of technology diffusion and org. learning |
| Communities of practice t. | Herzberg's two factor t. | Social interdependence t. | T. of work performance |
| Competitive strategy | Institutional t. | Social motivation t. | Time, interaction, and performance t. |
| Construal level t. | IS success model | Social presence t. | Trust t. |
| Constructive controversy t. | Knowledge-based t. of the firm | Socio-technical model | Value network t. |
| Contextual integrity t. | Path dependence | Speech-act t. | Whistleblowing t. |

## 4.4 What types of theories are used?

Of the 87 theories, 75 (87%) are middle-range theories (e.g. TAM). The remaining twelve are general theories (e.g. structuration theory). Process theories (39) were more common than variance theories (36) or taxonomies (12). We did not identify any empirical generalizations (see Section 6.2: Limitations).

In terms of Gregor's theory types, the most common type was Type II: theories for explaining (38), followed by Type IV: theories for explaining and predicting (35), and Type I: theories for analyzing (13). These numbers track well with those in the previous paragraph because most process theories explain, most variance theories explain and predict, and most taxonomies analyze. Again, this shows that theories for explaining, but not predicting, are more common in SE than generally assumed [cf. 58].

One Type V (for design and action) theory and no Type III (for predicting) were identified. This is not surprising since Types III and V appear less common in social science [26].

## 4.5 For which purposes are theories used?

We identified 198 individual theory *uses* (Fig. 2). Some articles use multiple theories or use one theory in multiple ways. The most common roles are *design* (53%) and *explanation* (16%). That is, SE research tends to use theories to inform the methodological design of a study or explain and analyze the results. Theories are used to design research protocols and task materials; to formulate hypotheses, research questions and design frameworks or models; and to develop questionnaires and other instruments.

## 4.6 Where in the papers are theories used?

Theories are mostly used in method (40%) and introduction (26%) sections; less often in results (19%) or discussion (15%) sections. This distribution roughly corresponds to the purposes for which the theories are used. That is, since theories are often used to design a study, motivate the research, or inform further theorizing, they tend to be used in the introduction and method parts of a paper.

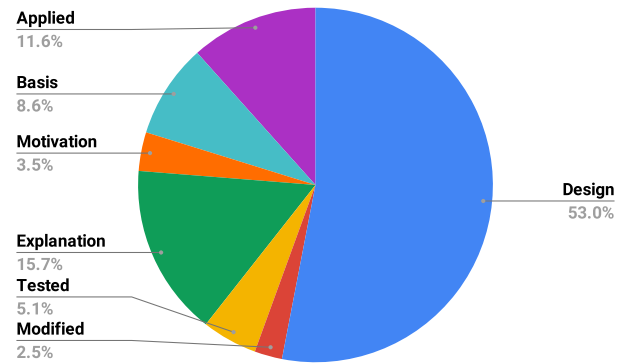

Figure 2: Distribution of theory roles

Applied 11.6%
Basis 8.6%
Motivation 3.5%
Explanation 15.7%
Tested 5.1%
Modified 2.5%
Design 53.0%

Theories are less often tested, or used to analyze results, which happens more in the results and discussion sections.

One might expect articles that test a social science theory to use that theory throughout, or at least in multiple sections. Rather, we found that the articles that tested a theory did so tangentially to their core purpose, such that the tested theory was mainly used in one section of the paper.

## 4.7 In what knowledge areas are theories used?

We attempted to map the topic of each article into the SWEBOK knowledge areas. The most common knowledge areas are professional practice (24) and models and methods (13). For example, Babar et al.'s [P68] study of software architecture evaluation in distributed versus co-located meetings exemplifies the professional practice area. Meanwhile, Yu and Petter's [P69] use of shared mental models theory to understand agile software development processes is a good example of the models and methods area. Many theories were also used in the context of software requirements (12), economics (9), and SE management (7). Theories were less used in the context of construction (4), SE process (4), testing (3), design (3), maintenance (3), and quality (1). No articles used theories in the context of other knowledge areas.

However, 19 articles did not map well into the SWEBOK knowledge areas. Of these, twelve can be best categorized as technology adoption (e.g. investigating technology adoption and acceptance in Botswana [P16]). Seven are about research methods (e.g. using the garbage can model to find a research method for studying open-source software [P2]).

Naturally, social science theories are used more in the primarily social knowledge areas than in the primarily technical knowledge areas. The more technical knowledge areas may use more mathematical and natural-science foundations, or may be less theoretical than the more social knowledge areas. That social theories are used so rarely in software design is puzzling.

Perhaps more concerning is that the SWEBOK knowledge area taxonomy appears deficient for classifying either SE research or SE practice. Much of the research we reviewed does not fit in any of the categories, and many common practices resist classification. How would one classify, for example, research on continuous integration practices?

## 4.8 Member checking results

Approximately 20% of the authors completed our questionnaire, leading to minor changes to the classification of some theories, mainly in how theories are used within the articles. A sample issue that was discussed upon analyzing the questionnaire was the origin of Adaptive Structuration Theory (MIS) compared to Structuration Theory (sociology). Benefits of using social science theories in SE research mentioned by respondents include:

- improving understanding of phenomena occurring in SE;
- theoretical grounding of their own research; and
- connecting technical and human aspects such as the way of thinking, teamwork, and failure.

Challenges mentioned include:

- understanding the theoretical background as social science theories are often more abstract;
- finding corresponding theories and mapping them to SE;
- theories that seem useful but lack necessary precision.

Some authors are sympathetic to theories, but find difficulty in publishing due to reviewers not understanding social science theories. One response mentioned the psychological uncertainty when using qualitative constructs and the challenge of accepting that "everyone does it differently".

## 4.9 Exploratory findings

Several findings beyond the original analysis questions emerged. This section briefly reports the more interesting incidental findings.

Software engineering is closely related to information systems, so we might expect a large overlap in theory use. However, only 28 (32%) of the theories used in our sample are listed in the *Theories Used in IS Research Wiki* or Dwivedi et al.'s [19] survey of theories commonly used in IS research.

Meanwhile, cross-referencing where theories are used with the role the theory plays highlights a few combinations:

- Design/method (29%) and design/introduction (23%) are the most common combinations. That is, the theory informs the design of the study, and this is explained in either the introduction or the method section. For example, Babar et al. [P68] formulate their hypothesis based on task-media fit in the methodology section. Marsan et al. extend concepts of institutional theory and rhetorical theory to propose a research question in their introduction [100].
- Explanation/discussion is also common (11%). For example, Hyrynsalmi et al. [102] use cultural lag theory to explain value creation mechanisms for mobile applications in their discussion section.
- Motivation/introduction, applied/method, and basis/method each make up another approximately 3%. The first is self explanatory. As an example of the second, Yu and Petter [P69] used shared-mental models theory to analyze different kinds of agile practices in their methods section. Moody [P1] introduces a design theory on the basis of various existing theories, as an example for the third.

The number of articles contributed by each journal varies greatly. TOSEM and ESE contributed only 274 and 642 papers respectively, while IST and JSS contributed 1419 and 2453. TSE was in the middle with 792. This basically reflects publication numbers—JSS and IST publish a lot more articles.

However, even after accounting for their relative contribution to the sample, journals vary significantly in their use of theories. Less than 1% of the TOSEM papers used theories, followed by TSE with 1.51%, JSS with 1.59%, ESE with 1.87%, and 2.60% IST papers. This suggests that either IST is most sympathetic to research using social science theories, or that researchers who use social science theories prefer submitting to IST.

We might expect more articles using social science theories in recent years, as SE has become more concerned with the human and social aspects of development. However, no particular trends are evident in the data overall, or within any specific journal.

## 5 MOST USED THEORIES

This section describes in more detail some of the most used theories in our sample. While space limitations prevent comprehensive descriptions of all 87 theories, elaborating a few prevalent ones should give the reader a better sense of the sort of social science theories being used in SE research.

### 5.1 The Technology Acceptance Model

The most used theory in this review is the technology acceptance model (TAM). TAM is a middle-range variance theory for explaining and predicting organizational adoption of new technologies [17]. TAM posits that an organization's intention to adopt a technology depends on perceived usefulness and perceived ease of use. It is based on Fishbein and Ajzen's theory of reasoned action [23].

TAM is arguably the most widely used theory in information system research. It has been extensively tested, refined and extended, most notably by the Unified Theory of Acceptance and Use of Technology (UTAUT; Fig. 3) [80].

However, TAM has also been widely criticized [12]. For example, Benbasat and Barki [5] argue that TAM "has created an illusion of progress in knowledge accumulation" and attempts to extend it have created "a state of theoretical chaos". More generally, no one ever argued that software should be useless and difficult to use. Pointing out that usefulness and usability are important is not terribly helpful without some theory of what makes a software product useful and usable.

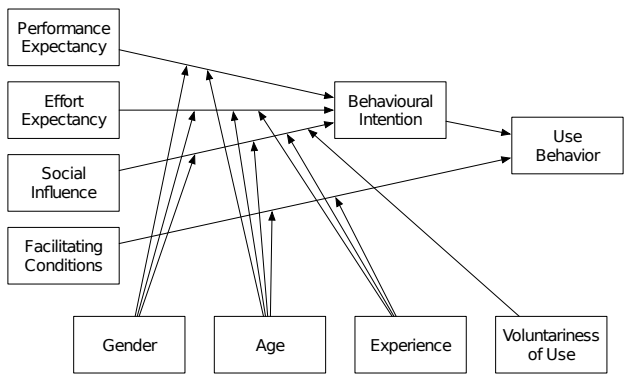

Figure 3: Unified theory of acceptance and use of technology [80]

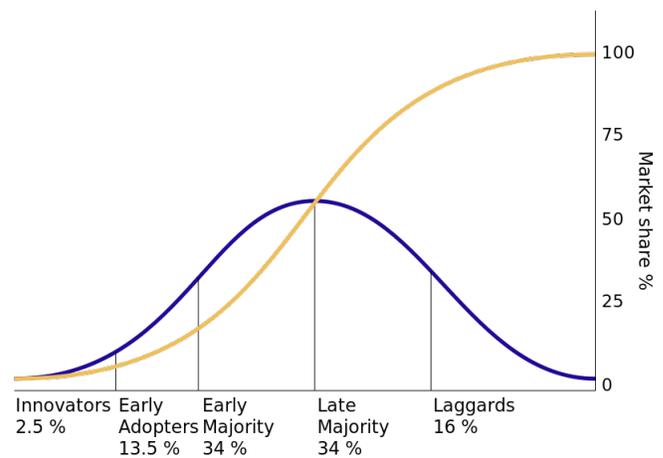

Figure 4: Diffusion of innovations (adapted from [65])

In twelve out of thirteen articles, TAM is used in the design role. For instance, Mohagheghi et al. [P4] employ TAM to construct an interview and survey about model-driven engineering.

### 5.2 Diffusion of Innovations

Diffusion of innovations is a middle-range process theory for explaining how innovations spread through communication channels over time within a social system [65]. It models diffusion as a kind of communication in which a new idea or product is shared among individuals in a social system.

The main elements of this theory are innovation, communication channels, time, and a social system. An innovation is simply an idea or product that is perceived as new by potential adopters. Communication is the process in which the participants in the social system share information to create consensus. Communication channels (including mass media and interpersonal communication) are the means by which an innovation is communicated among members of a social system. Meanwhile, "a social system is defined as a set of interrelated units that are engaged in joint problem solving to accomplish a common goal. The members or units of a social system may be individuals, informal groups, organizations, and/or subsystems" [65].

The theory posits that innovations diffuse in five stages [65]:

(1) **Knowledge.** An individual learns about an innovation.
(2) **Persuasion.** The individual forms an opinion (positive or negative) about the innovation.
(3) **Decision.** The individual decides whether to adopt or reject the innovation. This may include trying a new product or attending demonstrations.
(4) **Implementation.** The individual uses the innovation and judges its usefulness. If an individual decides to adopt an innovation, it is put to use during the implementation stage.

(5) **Confirmation.** The individual decides to retain or abandon the innovation. This stage can last indefinitely.

Diffusion of innovations further posits five types of adopters. From quickest to slowest to adopt, they are: innovators, early adopters, the early majority, the late majority, and laggards (Fig. 4).

Diffusion of innovations is not without its critics. For example, Lyytinen and Damsgaard [40] present six criticisms:

(1) "Technologies are not discrete packages."
(2) "Technologies do not diffuse in a homogenous and fixed social ether."
(3) Push and pull forces do not fully explain diffusion rates.
(4) The theory does not include all of the parameters that influence adoption decisions.
(5) "Diffusion does not necessarily traverse through distinct stages, which exhibit no feedback."
(6) "Time scales are not necessarily short and the history of decisions is not unimportant."

Diffusion of innovations is used in five papers and is therefore tied (with dual coding theory; next) for second most common theory.

### 5.3 Dual Coding Theory

Dual Coding Theory is an influential general theory for explaining memory and cognition [49]. It posits that people have two cognitive subsystems: one for words and one for images. Our minds therefore use two main kinds of representations or "codes:" analogue codes (images) and symbolic codes (words). The theory aims to explain how these two subsystems interact and affect cognition. Therefore, presenting the same idea using both words and images should help people remember, because the idea is encoded in both cognitive subsystems.

However, dual coding theory remains controversial. Many psychologists prefer the *common coding* theory of memory, which posits a single kind of representation and cognitive system [76].

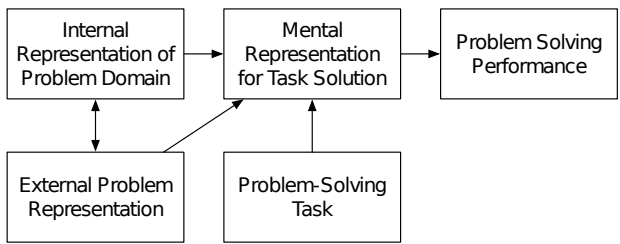

Figure 5: Cognitive fit theory (adapted from [67])

### 5.4 Coordination Theory

Coordination theory is a general theory for explaining "how the activities of separate actors can be coordinated" [41]. It includes a framework for analyzing coordination in complex processes, as well as a typology of dependencies and coordination mechanisms [42]. It posits that in an organization, some tasks depend on other tasks, which creates *coordination problems*. Actors solve these problems using *coordination mechanisms*. Coordination theory posits different kinds of coordination problems and mechanisms.

Coordination theory is used to design not only human organizations but also technologies (e.g. tools for facilitating coordination, distributed computer systems) [41].

### 5.5 Cognitive Fit Theory

Cognitive fit is a middle range, variance theory for explaining and predicting task performance. It "describes the relationships between graphical and tabular representations and the types of tasks they support" [81]. It posits that matching information representations to problem solving tasks will increase task performance (Fig. 5). For example, to navigate, we might want a map, but to calculate distance traveled, we might want a table showing distances between destinations.

Cognitive fit theory has evolved over time. For example, Samuel et al. [66] extend cognitive fit theory to introduce the dual-domain problem solving framework. While uncontroversial, cognitive fit is only one of multitudinous antecedents of task performance, and may be overwhelmed by other, more important factors.

## 6 DISCUSSION

### 6.1 Implications

Our analysis produced several surprising findings:

(1) Less than 2% of the articles use a social science theory in a substantive way.
(2) SE research rarely *tests* a social science theory, for example, to see if it actually applies to SE.
(3) Many relevant social sciences are not referenced at all (e.g. criminology).
(4) Process theories are used more often than variance theories, which shows that process theories are much more common in SE than generally assumed [cf. 58].

Intuitively, less than two percent of papers using social theories seems low. For comparison, at least two thirds of the papers in the June 2021 issue of *MIS Quarterly* appear to use one or more social theories. At least three possible explanations are evident:

(1) SE research primarily uses homegrown theories;
(2) SE research primarily uses natural science theories;
(3) SE research is primarily atheoretical.

Distinguishing between these three explanations would require a different kind of study, with a different sampling strategy. However, based on our experience with initiatives aimed at encouraging theory development and use, we suspect SE research is largely less concerned with theory (more atheoretical) than other similar fields. Compared to management, for example, SE researchers appear more concerned with making tools than theories.

Additionally, some of the theories that are being used are problematic. TAM, for example, is outdated [80], does not explain a core SE phenomenon, and has been widely criticized [12].

Broadly considered, the results of this study suggest that SE research is not making good use of social science theory. Failing to incorporate social theories is problematic in several ways.

Theories underpin almost all natural, social and applied sciences. Sound theoretical foundations are essential for maturing into a fully developed research discipline [70]. Without a collection of core theories, a discipline struggles to generate, accumulate and preserve knowledge [62, 68, 70].

Second, SE is intrinsically interdisciplinary—it intersects computer science, economics, management, mathematics, philosophy, psychology, and sociology. The social and technical aspects of creating software are inextricably entwined; they cannot be cleaved apart and understood independently. More than two percent of SE research needs to draw on social science theories to have any hope of understanding the full scope of software development.

In our experience, this inattention to social theory already harms SE research, practice, and education. Some SE research tends to oversimplify and over-rationalize social phenomena. For example, where an SE researcher asks "what is the problem the system should solve?" a sociologist sees multiple stakeholders, who do not agree on the nature of the problem or the aims of the system [11], jockeying for control of a project's agenda—there is no "the problem". Where a software professional sees "requirements elicitation", a psychologist sees a prospective user and an analyst analyst co-constructing preferences *specific to that moment and context* [37]—the outcomes are neither "requirements" nor "ellicited" [54]. Moreover, while students of other applied sciences learn to design innovative systems based on ill-defined opportunities, SE students primarily learn to construct routine, fully-specified systems, as if figuring out what to build is someone else's job [cf. 53].

Furthermore, infrequent use of social theory is surprising given two trends in SE research:

(1) The SE academic community is midway through a shift from rationalism (it works because it intuitively makes sense) to empiricism (it works because the results of this empirical study demonstrate that it works) [57].
(2) The SE community appears increasingly receptive to sociological research methods including grounded theory [73] and ethnography [83].

As the community demands more rigorous research, demands more empirical studies, and attempts more qualitative and social research, we expect researchers increasingly to draw on reference disciplines for both supporting foundations, evidence and methodological guidance.

## 6.2 Limitations

The results of this study should be weighed against several limitations. It is very likely that some articles in the sample were missed because they used a theory, but not the word "theory." We mitigated this threat by searching for a list of specific theories by name, but that list too is incomplete.

Similarly, our search process may not have been sensitive to empirical generalizations. These generalizations would not be listed in the IS Theory Wiki or or Dwivedi et al.'s [19] review, and would not normally have "theory" in their names, if they even have names. Therefore, the fact that we did not uncover any empirical generalizations from social science does not mean none were used.

The five journals and 13-year period we studied are not representative of all SE outlets across all time. We used purposive, search-based sampling [4] to select *good, influential* journals rather than random journals and focused on more recent years because we want to understand the *trajectory* of the field, which is mostly set by recent papers in top outlets.

Meanwhile, researchers do not agree on the definition of *theory*. Different researchers might have included or excluded different theories. We mitigate this threat by providing a complete list of the included theories (Section 4.2) and discussing edge cases (Section 3.3.1). We also tried searching for "model", but this produced too many false positives.

We validated our classifications through audits (i.e. the second and third authors extensively reviewing the classifications) and member checking (i.e., sending a questionnaire to primary study authors). This led to significant recategorization of Gregor's theory types and a handful of minor changes. However, different analysts may classify theories, usages and topic areas differently.

## 6.3 Areas for future research

SE researchers should consider using more social theories from reference disciplines. Using social theories more often, and more effectively, will help SE researchers produce more rigorous, nuanced, insightful research. Furthermore, SE researchers should move beyond using theories to motivate their papers or specify the research methodology. We need to *test* social theories in SE contexts and theorize about SE phenomena using concepts from the social sciences. For example, iteratively evaluating and adapting the theory of boundary objects [69] could help us better understand product backlogs, user stories, and other common non-code artifacts.

Similarly, social theories are often used to examine project management, requirements engineering, modeling, software development methods and technology adoption. SE researchers tend not to use them to examine designing, constructing, testing, evaluating and configuring software. However, the latter are socio-technical processes in which technical phenomena are entangled with social phenomena. Again, social theories should help. For example, numerous theories concerning power and influence could help explain interpersonal conflict during peer programming.

Additionally, SE researchers lean toward theories from psychology (to understand professionals) and management (to understand teams and organizations). We rarely draw on education (to understand professional development), criminology (to understand deviant behavior), health (to understand professionals' wellbeing), history (to understand the evolution of SE), or philosophy (to understand a great many things). Exploiting the vast array of relevant knowledge from myriad disciplines may help launch many fruitful research programs.

More specifically, this study investigates use of social science theories. Similar reviews of homegrown SE theories and natural science theories used in SE are needed.

## 6.4 Comparison to previous reviews

Several studies have examined the status of theory use in SE research. Hannay et al. [27] present a systematic literature review of theories used in software engineering experiments. Stol & Fitzgerald [70] argue that SE researchers often use "theory fragments" for analysis. Lim et al. [38] examine the theoretical basis of information systems research. Ralph et al. [59] describe seven theories that, they argue, ought to be core to SE research.

Like Hannay et al., we find that theories are mostly used to motivate and design studies; fewer articles test or modify theories. Like Lim et al., we find that TAM is widely-used, and that psychology and economics are popular reference disciplines. Like Ralph et al., we argue that SE researchers should use more social theory.

However, this review extends previous work in several ways. Stol & Fitzgerald [70] is more of a position paper with a limited review. Hannay et al. [27] only consider experiments, and their review is getting dated. Lim et al. [38] looked at a different field. Ralph et al. [59] recommend social theories rather than studying which are currently used. In summary, we extend these reviews by studying a larger, broader, more recent sample of primary studies, leading to somewhat different recommendations.

## 7 CONCLUSION

The presented study examined substantive uses of social science theories in more than 5500 articles published in five leading software engineering journals over a thirteen-year period. Despite several calls for more attention to theory in SE research, increasing attention to the social aspects of development, and increasing use of behavioral and qualitative research, *only about two percent of the reviewed articles use social science theories* in any substantial way. In other words, SE research tends to ignore relevant theories from social sciences.

When social science theories are used, they are most commonly:

- middle-range theories;
- process theories;
- theories for explaining (but not predicting);
- drawn from psychology or management;
- used in some way related to designing the study;
- used in the introduction or methods section of the paper;
- used in research concerning professional practice, technology adoption or models and methods of software development; and
- published in *Information and Software Technology*.

In contrast, SE research tends not to:
- test theories being used;
- use theories to analyze data;
- apply theories to software design, testing, evaluation or configuration;
- draw theories from relevant reference disciplines including health, criminology or history.

Applied sciences benefit from balancing more concrete techniques and more abstract theories. Techniques encapsulate practical, specific, contextualized, often short-term knowledge that practitioners can readily apply. Theories encapsulate more abstract, generalizable, long-term knowledge. They insulate a field against fads, guide research programs and help us understand novel situations.

For SE research to mature and improve, researchers therefore need to make better use of theory: social science theories to understand social phenomena, natural science theories to understand physical and mathematical phenomena, and more homegrown theories to understand SE-specific phenomena.

No one is saying that SE should abandon developing and evaluating new technologies or techniques. Rather, SE has twin aims of understanding software development and creating tools and techniques for developing software better. Our abilities to build better systems and to explain software development phenomena necessarily co-evolve. Many of our hardest problems lie at the intersection of social phenomena, technical phenomena, understanding, and creating. Only by accepting the fundamental interconnections among these four dimensions can we address our greatest challenges.

## DATA AVAILABILITY

The anonymized replication package includes the questionnaire template, the list of primary studies, and a spreadsheet containing all extracted data on articles and theories being used, as well as decision rules for inclusion and exclusion criteria. It is available at https://doi.org/10.5281/zenodo.6036076.